\documentclass[amsmath, amssymb, preprintnumbers, showpacs, showkeys, aps,prb,reprint]{revtex4-1}
\usepackage{graphicx}
\usepackage{color}
\usepackage{braket}
\usepackage{ulem}   
\usepackage{slashed}
\newcommand{\bs}[1]{{\boldsymbol{#1}}}

\begin{document}
\preprint{}

\title{M\"{o}bius molecules and fragile Mott insulators}
\author{Lukas Muechler$^1$}
\author{Joseph Maciejko$^{2,3}$}
\author{Titus Neupert$^3$}
\author{Roberto Car$^1$}
\affiliation{$^1$Department of Chemistry, Princeton University, Princeton, New
Jersey 08544, USA\\
$^2$Department of Physics, University of Alberta, Edmonton, Alberta T6G 2E1, Canada\\
$^3$Princeton Center for Theoretical Science, Princeton University, Princeton, New Jersey 08544, USA}

\date{\today}

\begin{abstract}
Motivated by the concept of M\"obius aromatics in organic chemistry, we extend the recently introduced concept of fragile Mott insulators (FMI) to ring-shaped molecules with repulsive Hubbard interactions threaded by a half-quantum of magnetic flux ($hc/2e$). In this context, a FMI is the insulating ground state of a finite-size molecule that cannot be adiabatically connected to a single Slater determinant, i.e., to a band insulator, provided that time-reversal and lattice translation symmetries are preserved. Based on exact numerical diagonalization for finite Hubbard interaction strength $U$ and existing Bethe-ansatz studies of the one-dimensional Hubbard model in the large-$U$ limit, we establish a duality between Hubbard molecules with $4n$ and $4n+2$ sites, with $n$ integer. A molecule with $4n$ sites is an FMI in the absence of flux but becomes a band insulator in the presence of a half-quantum of flux, while a molecule with $4n+2$ sites is a band insulator in the absence of flux but becomes an FMI in the presence of a half-quantum of flux. 
Including next-nearest-neighbor-hoppings gives rise to new FMI states that belong to multidimensional irreducible representations of the molecular point group, giving rise to a rich phase diagram.
\end{abstract}


\maketitle

\section{Introduction}

While the term \textit{strong correlations} is commonly used to describe a broad range of interacting systems, one typically considers a fermionic system to be strongly correlated if the quasiparticle picture breaks down, i.e., there is no continuous evolution between the noninteracting system and its interacting counterpart. However, strong correlations are not in one-to-one correspondence with strong interactions. On the one hand, a system can be strongly interacting (large Hubbard $U$ or Hund's coupling $J$) yet not be strongly correlated in the above sense, as commonly encountered in transition metal oxides.\cite{kuebler,exchangeeffect,dlmoxide,Moxquasipart,MnOlocalmoment,georges2013strong}
On the other hand, while in two and higher dimensions a Fermi liquid is stable against weak repulsive interactions, one-dimensional (1D) systems can exhibit a strongly correlated phase already at small values of $U$ and $J$, leading to a Luttinger liquid.\cite{luttliq} In quantum chemistry this distinction corresponds to the difference between dynamical and static correlation.
Dynamically correlated systems interact strongly, yet there are well defined quasiparticles and perturbative methods can be applied. This description breaks down for statically correlated systems, a classic example of which is the dissociation of the $H_2$ molecule beyond the Coulson-Fischer point.\cite{coulsonfischer} At this point, a single Slater determinant is not sufficient to describe the correct physics even qualitatively.

\begin{figure}[t]
 \centering
 \includegraphics[width=\columnwidth]{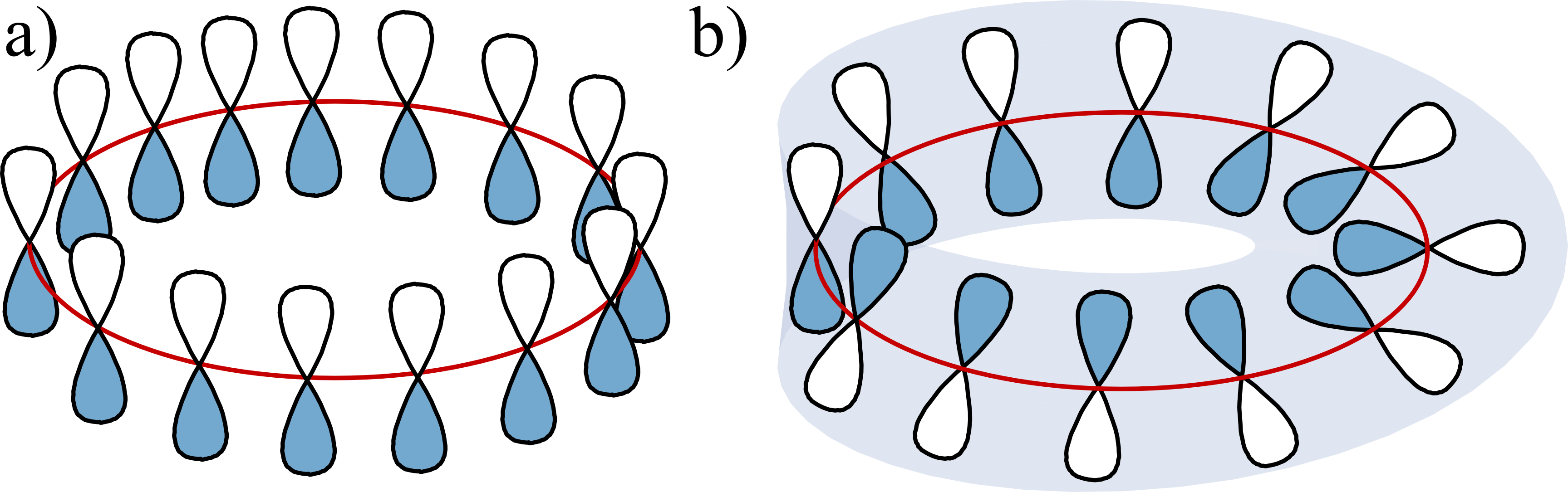}
 \caption{Orbital topology of (a) an aromatic molecule and (b) a M\"obius aromatic.}
\label{fig_1}
\end{figure}

The theoretical model most widely used to study correlation effects in fermionic systems is the Hubbard model. In particular, it is well known that the 1D Hubbard model exhibits distinct behavior for systems with $4n$ and $4n+2$ sites, where $n$ is an integer, when periodic boundary conditions (PBC) are assumed.
This difference in behavior is also well studied in organic chemistry. Molecules of the form C$_N$H$_N$ are called aromatic if $N=4n+2$ and anti-aromatic if $N=4n$. Aromatic molecules, such as benzene (C$_6$H$_6$), have a unique chemistry due to their chemical stability as well as a complex response to magnetic fields due to the aromatic ring current.\cite{ringcurr_3,*fowler2007aromaticity,*steiner2001four} Anti-aromatic compounds can be as stable as aromatic compounds if the topology of the orbital arrangement is that of a M\"obius band [Fig.~\ref{fig_1}(b)].\cite{Heilbronner} Remarkably, such M\"obius aromatics have been successfully synthesized.\cite{moebius1,*moebius2,*moebius3,*moebius4,*moebius5} 
The M\"obius topology of the orbital arrangement is equivalent to a ring with PBC but threaded by a half-quantum of magnetic flux $hc/2e$ (also known as a $\pi$ flux), or equivalently to a system with antiperiodic boundary conditions (aPBC) and zero flux.\cite{zoltan} \\

Perhaps surprisingly, the simplest member of the $4n$ family---the Hubbard square with $n=1$---forms an interesting strongly correlated state at half-filling, the fragile Mott insulator (FMI).\cite{steve_fragile} In general, a FMI is an insulator that cannot be adiabatically connected to a band insulator (BI) under the condition that time-reversal symmetry and certain point-group symmetries are preserved. The ground-state wave function of a BI is a single Slater determinant that must transform as the identity (trivial) representation of the point group, whereas the FMI is a correlated state whose ground-state wave function transforms as a nontrivial representation of the point group. For any $U > 0$, the ground state of the Hubbard square is unique and transforms as the $d_{x^2-y^2}$ representation of the $C_{4v}$ point group (i.e., the spatial symmetry group of the molecule as a whole) with a $C_4$ eigenvalue of $-1$.

In this paper, we use numerical and analytical methods to explore the interplay between interaction and correlation in more generic Hubbard molecules with time-reversal and point group symmetries. After a brief review of the concept of FMI (Sec.~II), we extend this concept to M\"{o}bius molecules (Sec.~III) and find both weakly correlated BI phases and strongly correlated FMI phases in two representative examples---molecules with $N=4$ and $N=6$ sites (Sec.~IV). Results in the general cases of $N=4n$ and $N=4n+2$ are then inferred from existing Bethe ansatz studies (Sec.~V). Adding a next-nearest-neighbor  (NNN) hopping to the $N=4$ and $N=6$ molecules, we find an even richer set of FMI phases, some corresponding to higher-dimensional irreducible representations of the molecular point group (Sec.~VI).

\section{Fragile Mott insulators}

In this section we give a brief review of the concept of FMIs.\cite{steve_fragile} We consider spinful fermions governed by a noninteracting Hamiltonian $\mathcal{H}_0$ that commutes with the antiunitary time-reversal symmetry operator $\mathcal{T}$ with $\mathcal{T}^2=-1$. The single-particle eigenstates of $\mathcal{H}_0$ are Kramers doublets 
$\ket{n}$ and $\mathcal{T} \ket{n} \equiv \ket{\tilde{n}}$ with 
$\braket{n|\tilde{n}} = 0$. In a second-quantized formulation where $c^\dagger_n$ creates a fermion in the single-particle state $|n\rangle$, a
BI is a state in which the members of a Kramers doublet are either both occupied or both unoccupied, 
\begin{equation}
\ket{\mathrm{BI}} = \prod_{n,\tilde{n} \in \mathrm{occ}} c^{\dag}_n c^{\dag}_{\tilde{n}} \ket{0},
\end{equation}
where $\ket{0}$ is the vacuum state with no fermions. We assume that $\mathcal{H}_0$ also possesses a unitary symmetry represented by the operator $\mathcal{R}$, i.e., $[\mathcal{H}_0,\mathcal{R}] = 0$. The single-particle states $\ket{n}$ can then be chosen to be eigenstates of the symmetry operator $\mathcal{R}$ with eigenvalues $\lambda_n$,
\begin{equation}
\mathcal{R}\ket{n} =  \lambda_n \ket{n}.
\end{equation}
If the unitary symmetry $\mathcal{R}$ is such that $\mathcal{R}^N = 1$ for some integer $N\geq 1$, the eigenvalues of $\mathcal{R}$ lie on the unit circle in the complex plane,
\begin{equation}
\lambda_n = e^{i2\pi\ell_n/N},\hspace{5mm}\ell_n =1,\ldots,N.
\end{equation}
Furthermore, we assume that the symmetry operation $\mathcal{R}$ commutes with time-reversal symmetry $[\mathcal{R},\mathcal{T}]=0$. In this case the $\mathcal{R}$ eigenvalues of the Kramers partners are related by complex conjugation $\lambda^{\ast}_n = \lambda_{\tilde{n}}$, since
\begin{equation}
\mathcal{R}\ket{\tilde{n}} = \mathcal{R}\mathcal{T}\ket{n} = \mathcal{T}\mathcal{R}\ket{n} = \mathcal{T}\lambda_n\ket{n} \\= \lambda^{\ast}_n \mathcal{T}\ket{n} = \lambda^{\ast}_n \ket{\tilde{n}},
\end{equation}
which implies that the band-insulator ground state $\ket{\mathrm{BI}}$ transforms trivially under the symmetry $\mathcal{R}$,
\begin{equation}\label{BItrans}
\mathcal{R}\ket{\mathrm{BI}} = \prod_{n,\tilde{n} \in \mathrm{occ}} \lambda_n \lambda_{\tilde{n}}  \ket{\mathrm{BI}} = \ket{\mathrm{BI}}.
\end{equation}
By contrast, a FMI is an insulator such that its ground state $\ket{\mathrm{FMI}}$ transforms nontrivially under $\mathcal{R}$,
\begin{align}\label{FMItrans}
\mathcal{R}\ket{\mathrm{FMI}}=\lambda\ket{\mathrm{FMI}},
\end{align}
with $\lambda\neq 1$. By virtue of Eq.~(\ref{BItrans}), a FMI must be a correlated state that cannot be described by a single Slater determinant. In the present context of spinful fermions hopping on a translationally invariant ring-shaped molecule with $N$ sites, $\mathcal{R}$ is the operator for a translation by one lattice site (which can also be considered as a $C_N$ rotation in the point group of the molecule as a whole).

\section{M\"obius molecules}
\label{sec:mobius}

We consider the following second-quantized Hubbard Hamiltonian to model
a ring molecule threaded by a magnetic flux $\Phi$,
\begin{align}\label{Hamiltonian}
\mathcal{H}(\Phi) &= -t \sum_{\sigma}\left( \sum^N_{j=1} e^{i\phi_j}c^{\dag}_{j\sigma} c_{j+1,\sigma}  + \mathrm{H.c.} \right)\nonumber\\
&\hspace{5mm}+ U \sum_{j=1}^N c^{\dag}_{j\uparrow} c_{j\uparrow} c^{\dag}_{j\downarrow} c_{j\downarrow},  
\end{align}
where $c_{j\sigma}^\dag$ ($c_{j\sigma}$) creates (annihilates) an electron of spin $\sigma$ on site $j$ and we define $c_{N+1,\sigma}\equiv c_{1\sigma}$, which corresponds to PBC, $t>0$ is the hopping amplitude, and $U>0$ is the strength of the repulsive interaction. The total flux threading the ring is $\Phi = \sum_j \phi_j$, and we fix the total electron number to be $N$, which is half filling. 

All physical observables such as the total flux $\Phi$ are invariant under a 
local $U(1)$ gauge transformation
\begin{equation}
c^{\dag}_{j\sigma}\rightarrow e^{i\alpha_j}c^{\dag}_{j\sigma},
\qquad \sigma=\uparrow\downarrow.
\end{equation}
In contrast, the individual phases $\phi_j$ that appear in the hopping matrix elements of $\mathcal{H}(\Phi)$ are not invariant under this gauge transformation. While our results will not depend on the choice of gauge, 
we will work in the uniform gauge  $\phi_j=\Phi/N\equiv\phi$ from here on to make the derivation more transparent. We denote by $\mathcal{H}_{\mathrm{uni}}(\Phi)$ the Hamiltonian in this uniform gauge.

The Hamiltonian $\mathcal{H}(\Phi)$ possesses a family of translational symmetries labeled by $\varphi\in[0,2\pi)$, that are represented by $\mathcal{R}_{\varphi}$ in the uniform gauge Hamiltonian $\mathcal{H}_{\mathrm{uni}}(\Phi)$ with
\begin{equation}\label{Rvarphi}
\mathcal{R}_{\varphi} c^{\dag}_{j\sigma}\mathcal{R}_{\varphi}^{-1}= e^{-i\varphi} c^{\dag}_{j+1,\sigma}.
\end{equation}
For general $\Phi$, the Hamiltonian $\mathcal{H}(\Phi)$ is not time-reversal symmetric.
Only for the special values $\Phi=0,\pi$, corresponding to the ring and the M\"obius molecule, is it possible to define a time-reversal symmetry. 
For $\Phi=0$, time-reversal symmetry is represented by $\mathcal{T}=i\sigma_2\mathcal{K}$, where $\mathcal{K}$ stands for complex conjugation and $\sigma_2$ is the second Pauli matrix acting on the spin index $\sigma$. By contrast, for $\Phi=\pi$ time-reversal symmetry is represented by $\widetilde{\mathcal{T}}=U_{\pi} \mathcal{T}$ where $U_\pi$ is a unitary operator defined by
\begin{equation}
U_{\pi}c^{\dag}_{j\sigma}U_{\pi}^{-1} = e^{-i2\pi j/N} c^{\dag}_{j\sigma} .
\end{equation}
Of the family of translational symmetries $ \mathcal{R}_{\varphi}$, only $\mathcal{R}_{\phi}$ commutes with $U_\pi$, while for instance $[\mathcal{R}_0,\tilde{\mathcal{T}}] = e^{i 2 \pi /N}$.
For that reason, we will focus on the translational symmetry $\mathcal{R}_{\phi}$ from here on, because the proof of Eq.~(\ref{BItrans}) relied on the assumption that time-reversal and lattice symmetries commute.

Consider now the Hamiltonian $\mathcal{H}_{\mathrm{uni}}(\pi)$ in the noninteracting limit $U=0$. The $\mathcal{R}_{\phi}$ eigenvalues $\tilde{\lambda}_{n_{\pi}}$ of the single-particle eigenstates $\ket{n_{\pi}}$ lie on the unit circle with
\begin{equation}
\tilde{\lambda}_{n_{\pi}} = e^{i2\pi \left(\ell_{n_{\pi}} + 1/2 \right)/N},\hspace{5mm} \ell_{n_{\pi}} =1,\ldots,N,
 \end{equation}
because $\mathcal{R}_{\phi}^N =e^{-i\Phi}= -1$. Furthermore, because $[\mathcal{R}_{\phi},\tilde{\mathcal{T}}] = 0$ the eigenvalues satisfy 
$\tilde{\lambda}^{\ast}_{n_{\pi}} = \tilde{\lambda}_{\tilde{n}_{\pi}}$,
\begin{align}
\mathcal{R}_{\phi}\ket{\tilde{n}_{\pi}} & = \mathcal{R}_{\phi} \tilde{\mathcal{T}}\ket{n_{\pi}} 
= \tilde{\mathcal{T}}\mathcal{R}_{\phi}\ket{n_{\pi}} 
= \tilde{\mathcal{T}}\tilde{\lambda}_{n_{\pi}}\ket{n_{\pi}} \nonumber\\
& = \tilde{\lambda}^{\ast}_{n_{\pi}} \tilde{\mathcal{T}}\ket{n_{\pi}} 
= \tilde{\lambda}^{\ast}_{n_{\pi}} \ket{\tilde{n}_{\pi}}.
\end{align}
Thus a band-insulator ground state at $\Phi=\pi$,
\begin{align}
\ket{\mathrm{BI}_{\pi}} = \prod_{n_{\pi},\tilde{n}_{\pi} \in \mathrm{occ}} c^{\dag}_{n_{\pi}} c^{\dag}_{\tilde{n}_{\pi}} \ket{0},
\end{align}
transforms trivially under the translation operator $\mathcal{R}_\phi$,
\begin{equation}
\mathcal{R}_{\phi} \ket{\mathrm{BI}_{\pi}} = \prod_{n_{\pi},\tilde{n}_{\pi} \in \mathrm{occ}} \tilde{\lambda}_{n_{\pi}} \tilde{\lambda}_{\tilde{n}_{\pi}}  \ket{\mathrm{BI}_\pi} = \ket{\mathrm{BI}_\pi},
\end{equation}
which is the $\pi$-flux analog of Eq.~(\ref{BItrans}). By analogy with Eq.~(\ref{FMItrans}), this allows us to extend the concept of FMI to M\"obius molecules. We will say that a M\"obius molecule has an FMI ground state $\ket{\mathrm{FMI}_\pi}$ if it transforms nontrivially under $\mathcal{R}_{\phi}$,
\begin{align}
\mathcal{R}_{\phi}\ket{\mathrm{FMI}_\pi}=\tilde{\lambda}
\ket{\mathrm{FMI}_\pi},
\end{align}
with $\tilde{\lambda}\neq 1$.

\section{Molecules with 4 and 6 sites}

\begin{figure}[t]
 \includegraphics[width=0.95\columnwidth]{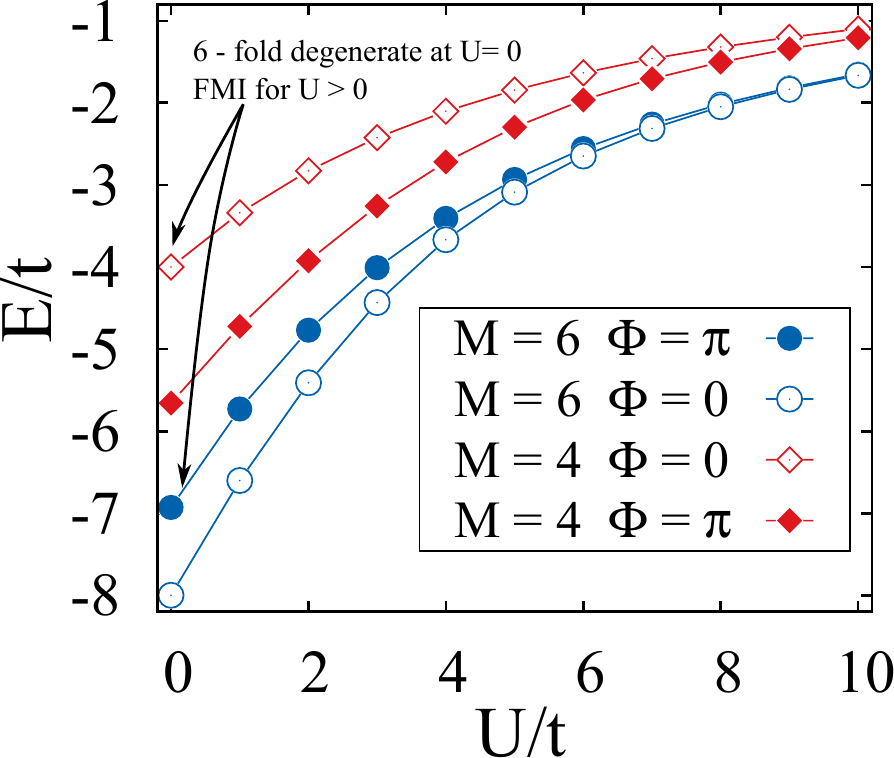}
\caption{Ground state energy of the 4-site (red diamonds) and 6-site (blue circles) half-filled Hubbard model with zero flux (open markers) and $\pi$ flux (filled markers), obtained by ED. The BIs are lower in energy than the FMIs.}
\label{fig_2}
\end{figure}
\begin{figure}[t]
 \centering
 \includegraphics[width=\columnwidth]{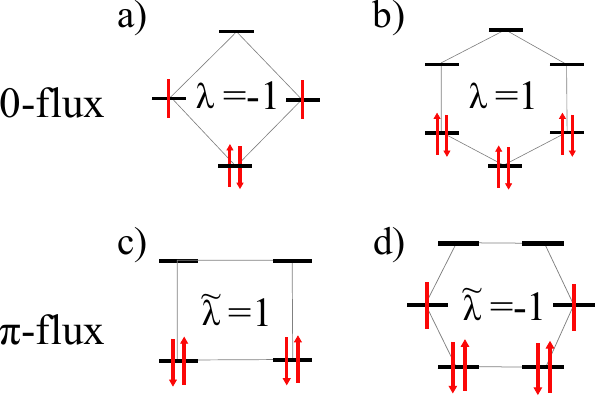}
 \caption{Single-particle energy levels for the 4- and 6-site rings at $U=0$. The insets give the $U>0$ ground-state eigenvalues $\lambda$ and $\tilde{\lambda}$ of the translation operators $\mathcal{R}_{0}$ and $\mathcal{R}_{\phi}$, respectively.}
\label{fig_3}
\end{figure}

As an illustration of the concepts presented above, and before discussing the general case with $N$ sites, we perform exact numerical diagonalization (ED) studies of the Hamiltonian (\ref{Hamiltonian}) for $N=4$ and $N=6$, and with $\Phi=0$ and $\Phi=\pi$. The low-energy spectrum (Fig.~\ref{fig_3}) allows us to determine whether the system is a metal or an insulator, and an explicit computation of the ground-state eigenvalue of the appropriate translation operator allows us to determine whether the system is a BI or a FMI. As mentioned previously, the ground state of the 4-site Hubbard model with zero flux and $U>0$ is a FMI with an $\mathcal{R}_{\phi}\,(=\mathcal{R}_{0})$ eigenvalue of $-1$.\cite{steve_fragile} At $\Phi=\pi$ however, the $\mathcal{R}_{\phi}\,(=\mathcal{R}_{\pi/N})$ eigenvalue is $+1$ for all $U\geq 0$ and thus the ground state is a BI. Although strictly speaking the theorem (\ref{BItrans}) assumed noninteracting electrons, we find no level crossing as a function of $U$ between the ground and first excited state, and the $\mathcal{R}_{\phi}$ eigenvalue of the ground state does not change. Therefore, to be more precise, one should say that the ground state at $U>0$ is adiabatically connected to the BI at $U=0$. On the other hand, we find that a ring with 6 sites at zero flux is a BI with a $\mathcal{R}_{\phi}$ eigenvalue of $+1$ for all $U\geq 0$. With a $\pi$ flux the situation reverses and the $\mathcal{R}_{\phi}$ eigenvalue is $-1$ for $U>0$, so the ground state is a FMI.

This behavior can can be understood qualitatively from the $U=0$ electronic structure. For 4 sites and 0 flux there are two degenerate single-particle states at the Fermi level, so the ground state cannot be a BI at half filling [Fig.~\ref{fig_2}(a)]. This is the molecular analog of a metallic state that turns into a Mott insulating state for $U >0$. For 6 sites, all single-particle states can be completely filled [Fig.~\ref{fig_2}(b)]. This is the molecular analog of a BI. In this case, interactions effects at small $U>0$ can be treated perturbatively and do not destabilize the state because of the single-particle gap. For a $\pi$ flux, the single-particle states for 4 sites can all be filled to give a band insulating state [Fig.~\ref{fig_2}(c)]. However, the 6-site system has degenerate single-particle states at the Fermi level and the situation is reversed with respect to the case of zero flux [Fig.~\ref{fig_2}(d)]. At $U=0$ the ground states of the BIs are unique, while the metallic states are six-fold degenerate. For both 4 and 6 sites, the ground states for $U>0$ are unique for both fluxes (Fig.~\ref{fig_3}).\cite{Lieb_peierls,nakano2000flux} At large $U$, the gap between ground and first excited state becomes very small. This can be understood from the fact that the large-$U$ limit of the Hubbard model is a Heisenberg model with exchange constant $J=4t^2/U$ that defines the energy scale.

In the next section, we generalize these results to molecules with any even number of sites $N\geq 4$. In analogy to M\"obius aromatics, we now prove that rings with $4n$ sites at zero flux and rings with $4n+2$ sites at $\pi$ flux are FMIs, since there will always be degenerate states at the Fermi level for $U=0$ in analogy to the 4- and 6-site rings discussed above. 

\section{Bethe ansatz and $4n$ vs $4n+2$}

The translations $\mathcal{R}_{\phi}$ are generated by the total gauge-invariant momentum,
\begin{align}\label{GImomentum}
P = \sum_{k\sigma} \left(k - \frac{\Phi}{N}\right) c^{\dag}_{k\sigma} c_{k\sigma},
\end{align}
through $\mathcal{R}_{\phi}=e^{iP}$, where the sum over $k$ is over all momenta in the first Brillouin zone $-\pi<k\leq\pi$. At half filling, $\sum_{k\sigma}c^\dag_{k\sigma}c_{k\sigma}=N$ and we have
\begin{align}
P = \sum_{k\sigma}k c^{\dag}_{k\sigma} c_{k\sigma}-\Phi,
\end{align}
i.e., the total gauge-invariant momentum is obtained by simply shifting the total canonical (non-gauge-invariant) momentum by the total flux $\Phi$. In the $\pi$-flux case, $\mathcal{R}_{\phi}$ corresponds to a translation generated by $P = \sum_{k\sigma}k c^{\dag}_{k\sigma} c_{k\sigma}-\pi$ (see Appendix~\ref{sec:app} for details).
Based on the $U \rightarrow \infty$ Bethe ansatz solution of the 1D Hubbard model\cite{Lieb_uniqueness,*thesis_uinf,*uinf_article} and due to the fact that the ground states are unique for all $U$ without any level crossing, we can prove that rings with $4n+2$ sites and a $\pi$ flux are FMIs characterized by a nontrivial eigenvalue of $e^{iP}$, which corresponds to a finite total momentum.  Rings with $4n$ sites are FMIs that become BIs upon inserting a $\pi$ flux.

For $U \rightarrow \infty$ at half filling and PBC, the momenta obtained from the Bethe ansatz are given by $k_j = 2\pi I_j/N$, where the $I_j$ are half-odd integers ($j - \frac{N+1}{2}$) for $N = 4n+2$ and integers ($ j - \frac{N}{2}$) for $N=4n$.\cite{Betheapbc}
The total momentum $\sum_j 2\pi I_j/N$ is thus zero for $N = 4n+2$ and $\pi$ for  $N = 4n$. For aPBC in the Bethe ansatz, which corresponds to a $\pi$ flux, the $I_j$ are integers for $N = 4n+2$ and half-odd integers for $N = 4n$, thus the total momentum is shifted by $\pi$.

\section{Next-nearest-neighbor hopping}

\begin{figure}[t]
 \includegraphics[width=0.90\columnwidth]{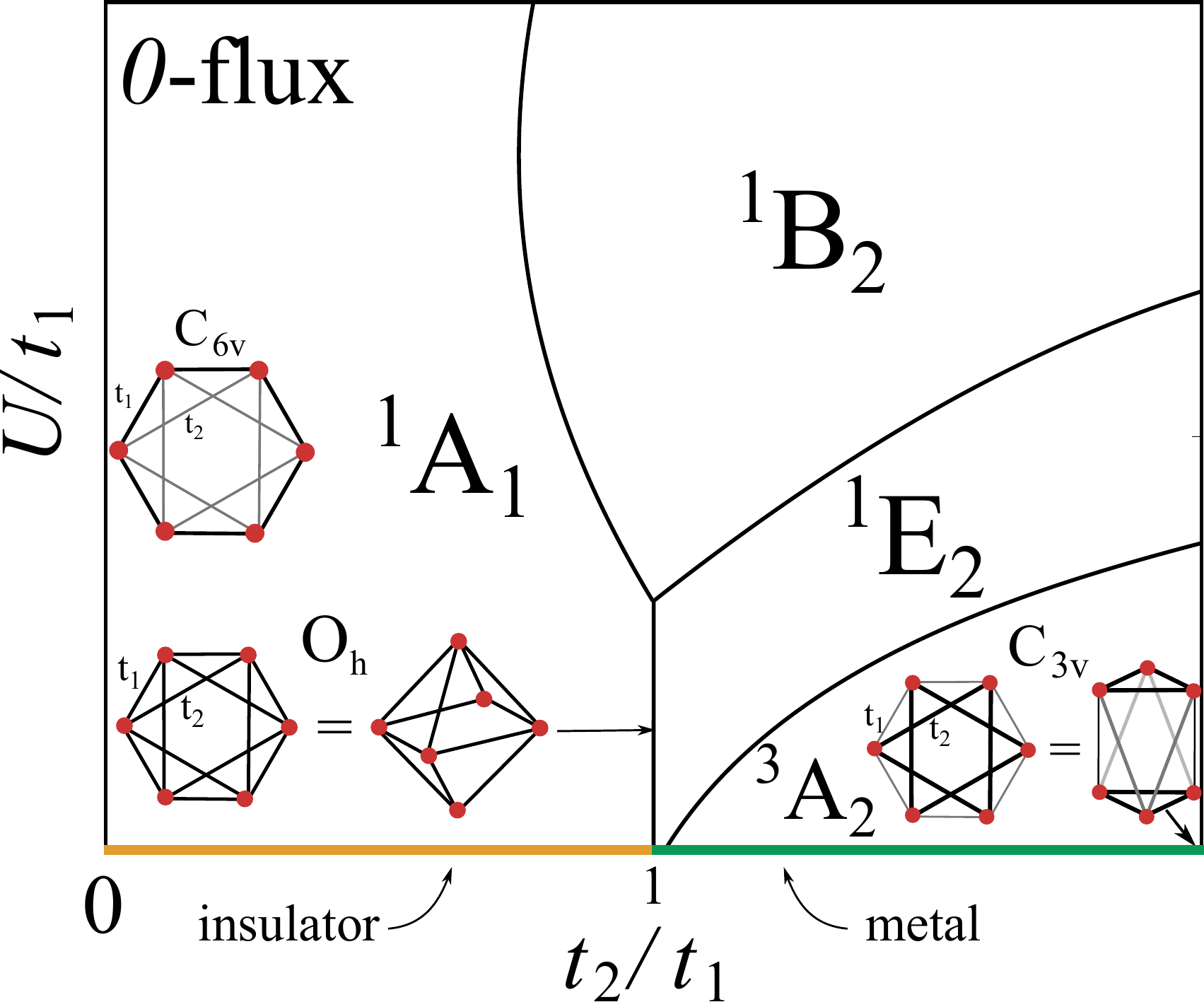}
\caption{Phase diagram for $N = 6$ at 0-flux with nearest-neighbor hopping $t_1$ and NNN hopping $t_2$, obtained by ED. Ground states are labeled by a symbol $^{2S+1}\Gamma$ where $\Gamma$ denotes the irreducible representation of the molecular point group $C_{6v}$ according to which the ground state transforms and $S$ is the total spin in the ground state.} 
\label{fig_4}
\end{figure}

Since we have considered a Hamiltonian with only nearest-neighbor hoppings so far, the stability of the FMI phases at $\pi$ flux with respect to NNN hoppings $t_2$ is an important question, as such terms are present in all realistic materials.
We consider a nnn hopping term of the form
\begin{align}
\mathcal{H}_{\mathrm{nnn}} = -t_2 \sum_{\sigma}\left( \sum^N_{j=1} e^{i2\phi_j}c^{\dag}_{j\sigma} c_{j+2,\sigma}  + \mathrm{H.c.} \right),
\end{align}
that preserves all the symmetries of the systems considered before if $\phi_j$ equals $0$ or $\pi/N$. Varying $t_2$ allows us to explore the relationship between the geometry of the molecule and its electronic structure. For example, in the 4-site model at zero flux, there is a phase transition at a critical value of the NNN hopping $t_2 = t_1$ where the system acquires an enhanced tetrahedral symmetry.\cite{steve_fragile}
Choosing the $N =6$ case as an example, we focus on the evolution of the single-particle levels at $U=0$, from which the behavior at $ U > 0$ can be understood.

The single-particle energies are given by
\begin{align}
\varepsilon(k) =- 2t_1 \cos(k_j) - 2 t_2 \cos(2k_j),
\end{align}
where $k_j= 2\pi j/N$ for 0 flux and $k_j= 2\pi(j+1/2)/N $ for $\pi$ flux where $j= -3,-2,\ldots,2$. For 0 flux and $t_2 = 0$, there are 6 single-particle levels (including spin) that can be completely filled and the system is a BI [Fig.~\ref{fig_2}(b)]. To introduce electronic frustration at 0 flux, the levels at $k = \pi$ and $ k = \pm\pi/3$ must cross, which happens at $t_2 = t_1$ where the system acquires an enhanced octahedral symmetry. 

Figure~\ref{fig_4} shows the phase diagram of the 6-site molecule at zero flux and $U > 0$ calculated by ED, with nnn hopping. Ground states are labeled by a symbol $^{2S+1}\Gamma$ where $\Gamma$ denotes the irreducible representation of the molecular point group $C_{6v}$ according to which the ground state transforms and $S$ is the total spin in the ground state. The BI phase $^1A_1$ is the ground state until $t_2 = t_1$ and for a large range of values of $U$.
When the symmetry of the problem is close to octahedral, three nontrivial correlated phases emerge. The $^1B_2$ phase is a unique FMI ground state that occurs at large values of $U$, whereas the $^3A_2$ phase
is a spin-triplet FMI state that occurs for $t_2 > t_1$. In the limit $t_2/t_1\rightarrow\infty$ the problem reduces to two decoupled staggered triangles. The intermediate $^1E_2$ phase is a doubly degenerate FMI with total momentum $P = \pm 2\pi/3$.
For $U=0$, the ground state is unique for $t_1 > t_2$. At $t_1=t_2$ the ground state is 15-fold degenerate and 6-fold degenerate for $t_2 > t_1$.
\\

\begin{figure}[t]
 \includegraphics[width=0.90\columnwidth]{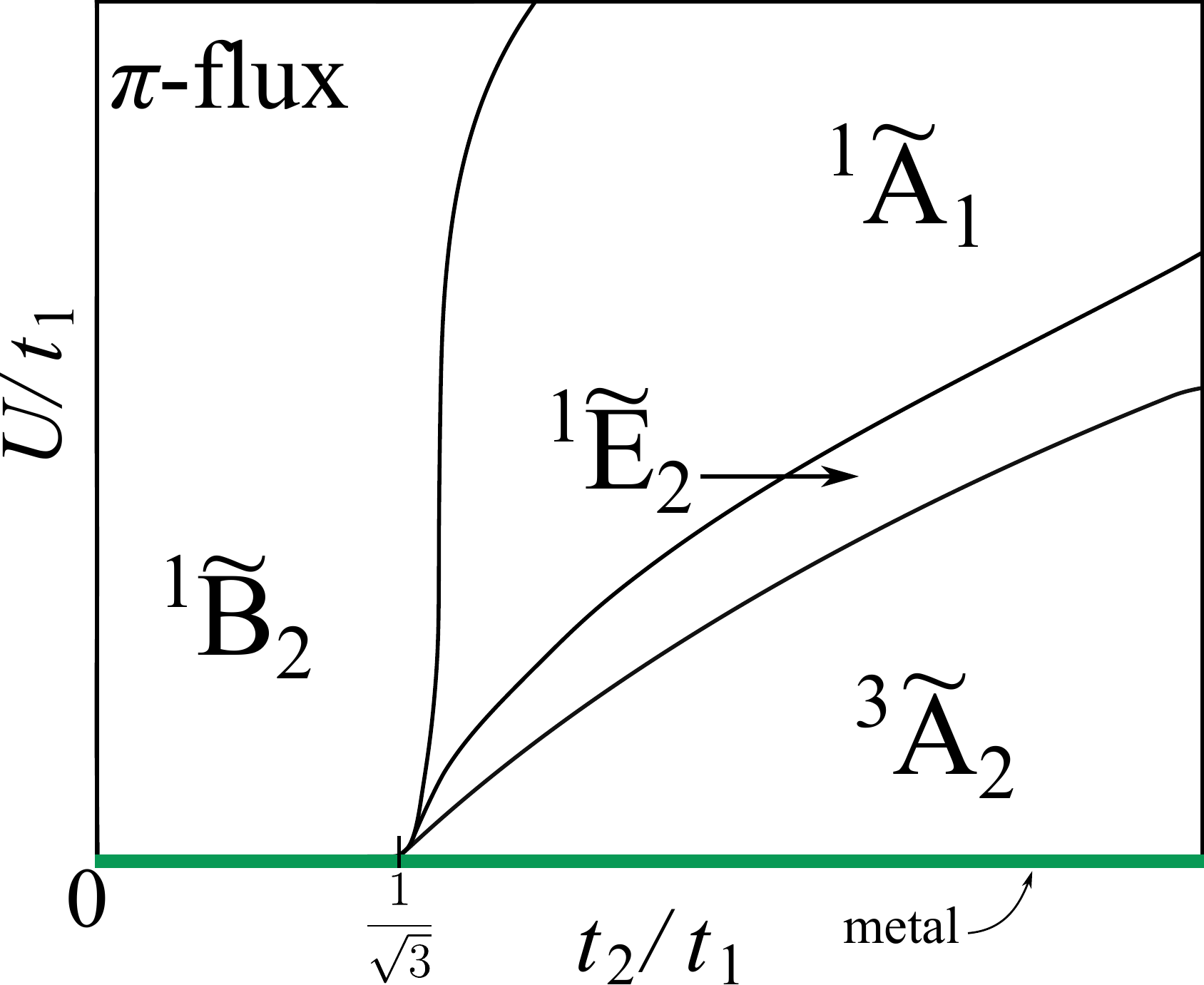}
\caption{Phase diagram for $N = 6$ at $\pi$-flux with nearest-neighbor hopping $t_1$ and NNN hopping $t_2$, obtained by ED. Ground states are labeled by a symbol $^{2S+1}\widetilde{\Gamma}$ where $\widetilde{\Gamma}$ denotes the irreducible representation of the molecular point group $C_{6v}$ according to which the ground state transforms and $S$ is the total spin in the ground state. $\widetilde{\Gamma}$ differs from $\Gamma$ in the definition of the translation operator ($\mathcal{R}_{\pi/N}$ for $\pi$ flux and $\mathcal{R}_0$ for zero flux, see Sec.~\ref{sec:mobius}).}
\label{fig_5}
\end{figure}

Figure~\ref{fig_5} shows the phase diagram for the same system but with a $\pi$ flux. At $t_2 = 0$, the many-body ground state is 6-fold degenerate at $U = 0$. There is a crossing of the doubly degenerate single-particle levels at $k_j = \pm 5\pi/6 $ and $k_j = \pm \pi/2 $  for $t_2/t_1 = 1/\sqrt{3}$. As in the absence of nnn hopping, the $\pi$ flux interchanges the FMI ($^1\widetilde{B}_2$) and BI ($^1\widetilde{A}_1$) phases with respect to the zero flux case. Around the level crossing the two other phases $^1\widetilde{E}_2$ and $^3\widetilde{A}_2$ emerge, similarly to the zero flux case. We use $\widetilde{\Gamma}$ instead of $\Gamma$ to denote the irreducible point group representations in the $\pi$ flux case simply to indicate that the translation operator or, alternatively, $C_N$ rotation operator should be taken as $\mathcal{R}_{\pi/N}$ for a $\pi$ flux, while it is $\mathcal{R}_0$ for zero flux (see Sec.~\ref{sec:mobius}).
The ground state at $U = 0$ is 6-fold degenerate for $ t_2/t_1 < 1/\sqrt{3}$ and $ t_2/t_1 > 1/\sqrt{3}$ . At the level crossing the ground-state degeneracy is 28.

\section{Conclusion}

We have shown that there is a duality between Hubbard rings with $N = 4n$ and $N = 4n+2 $ sites, which can be tuned by threading a magnetic flux through the ring. For zero flux, $4n$-membered rings are FMIs that cannot be adiabatically transformed into BIs if time-reversal symmetry and the molecular point group symmetry are preserved. Rings with $4n+2$ sites and a $\pi$ flux are also FMIs. All these FMI states do not break any symmetries and are characterized by a nonlocal order parameter, the total gauge-invariant momentum (\ref{GImomentum}). This order parameter is nonlocal because the lattice Fourier transform of $k$, as opposed to that of the periodic function $\sin k$, has a slow power-law decay $\sim 1/x$ in real space. The FMI is an example of non-fractionalized featureless Mott insulator protected by lattice symmetries.\cite{parameswaran2013,kimchi2013} In the $N \rightarrow \infty$ limit, the BI and the FMI become degenerate, since their distinction is due to boundary conditions (and the half-filled 1D Hubbard model has a gapless spin sector in the thermodynamic limit\cite{Giamarchi}).

From an organic chemistry point of view, the FMIs considered here are anti-aromatic molecules, whereas BIs are aromatic molecules. Anti-aromatic molecules typically appear as transition states, since BIs are lower in energy. A BI state is usually obtained by breaking the symmetry of the molecule through a spontaneous structural distortion. For example, the anti-aromatic molecule cyclooctatetraene (C$_8$H$_8$) cannot be isolated in a $D_{8h}$ symmetric structure since it adopts a tub configuration with lower $D_{2d}$ symmetry.\cite{cot1,cot2} Therefore, our findings appear to be related to the Woodward-Hoffmann rules.\cite{hoffmann1965selection,*woodward1969conservation} These rules only allow pericyclic reactions where the transition states are BI states, whereas reactions with an FMI as transition state are forbidden.

More broadly, our work illustrates the complexity of the relationship between interaction and correlation in fermionic systems. This is made most apparent by considering nnn hopping as a tuning parameter in our models, which mimics the effects of the geometric structure of the molecule on electronic properties. Weakly correlated BI phases do occur at small $U$ (e.g., the $^1A_1$ phase in Fig.~\ref{fig_4}) as one would expect, but they also occur at strong $U$ (e.g., the $^1\widetilde{A}_1$ phase in Fig.~\ref{fig_5}). Conversely, strongly correlated FMI phases do occur at large $U$ (e.g., the $^1B_2$ phase in Fig.~\ref{fig_4}) as one would expect, but they can also occur at small $U$ (e.g., the $^3A_2$ phase in Fig.~\ref{fig_4} and the $^1\widetilde{B}_2$ and $^3\widetilde{A}_2$ phases in Fig.~\ref{fig_5}). For small $t_2$, $U$ is not the deciding parameter in the phase diagram because it is the geometry that gives rise to electronic frustration. For large enough $t_2$ however, the precise value of $U$ plays an important role in determining the phase diagram.

Experimentally, it is very challenging to discriminate the FMI and BI phase directly, for their order parameters are hard to access and the measurement itself may not break the rotational symmetry explicitly. However, indirect evidence for the quantum phase transitions between FMI and BI could be obtained from spectroscopic measurements. For example, one could envision scanning tunneling microscopy experiments on molecules that are deposited on a solid substrate. \cite{gomes2012designer,pitters2011tunnel} If that substrate is a type-II superconductor, it is even conceivable that a $\pi$ flux can be trapped at the center of an arrangement of molecules in order to explore the phase diagrams presented in this work.

\bigskip 
\begin{acknowledgments}
L.M. would like to thank Claudia Felser and Shoucheng Zhang for important comments and discussions in the early stages of this work. We thank Steven Kivelson for providing useful references as well as Elliott H. Lieb and Zolt\'an G. Soos for stimulating discussions. This work was supported in part by the Simons Foundation (J.M.), the Natural Sciences Engineering Research Council (NSERC) of Canada (J.M.), the
DARPA grant SPAWARSYSCEN Pacific N66001-11-1-4110 (T.N.) as well as the Department of Energy grant DE-FG02-05ER46201 (L.M. and R.C.).
\end{acknowledgments}


\appendix
\section{Total momentum as generator of translations}
\label{sec:app}

To find the explicit form of the momentum operator, i.e., the generator of translations, we first consider the ansatz
\begin{align}
\mathcal{R} = e^{iP}=\exp\left(\sum_{nm} A_{nm} c^{\dag}_n c_m\right),
\end{align}
for the translation operator at zero flux $\mathcal{R}\equiv\mathcal{R}_0$ [see Eq.~(\ref{Rvarphi})], where $n,m=1,\ldots,N$ are lattice site indices. Using the Baker-Campbell-Hausdorff formula  
\begin{align}
e^{X}Y e^{-X} &=Y+\left[X,Y\right]+\frac{1}{2!}[X,[X,Y]]\nonumber\\
&\hspace{5mm}+\frac{1}{3!}[X,[X,[X,Y]]]+\ldots, 
\end{align}
where $X =  \sum_{nm} A_{nm} c^{\dag}_n c_m$ and $Y= c^{\dag}_l$, as well as the commutator
\begin{align}
[c^{\dag}_n c_m, c^{\dag}_l] = \delta_{ml} c^{\dag}_n,
\end{align}
we find
\begin{align}
e^{X} c^{\dag}_l  e^{-X} &= c^{\dag}_l + \sum_{n} A_{nl} c^{\dag}_n + \frac{1}{2!} \sum_{n} A^2_{nl} c^{\dag}_n
+ \frac{1}{3!} \sum_{n} A^3_{nl} c^{\dag}_n \nonumber\\
&\hspace{5mm}+ \ldots\nonumber \\
&= \left[\exp(\bs{A}^T)\bs{c}^{\dag}\right]_l,
\end{align}
where $\bs{A}$ is the $N\times N$ matrix with elements $A_{nm}$ and $\bs{c}^\dag$ is a $N\times 1$ column vector with elements $c^\dag_l$. The action of the translation should be $\left[\exp(\bs{A}^T)\bs{c}^{\dag}\right]_l= c^{\dag}_{l+1}$, which can be represented as $\left[\bs{T}\bs{c}^{\dag}\right]_l = c^{\dag}_{l+1}$ where $\bs{T}$ is a translation matrix. This implies that $\bs{A}^T = \ln\bs{T}$. PBC imply that $\bs{T}^N=1$, hence the eigenvalues of $\bs{T}$ are of the form $\lambda_n = e^{i2\pi \ell_n/N}$ with $\ell_n=1,\dots,N$. Furthermore, $\bs{T}$ is a diagonalizable matrix and its matrix logarithm is given by $\ln \bs{T} = \bs{V} (\ln\bs{T}_{\mathrm{diag}}) \bs{V}^{-1}$. Because $\bs{T}$ is a circulant matrix, the matrix of eigenvectors $\bs{V}$ is given by the kernel of the discrete Fourier transformation,
\begin{align}
(\ln\bs{T}_{\mathrm{diag}})_{kl} &=   2\pi i\left(\frac{\ell_k}{N}+\left[\frac{1}{2}-\frac{\ell_k}{N} \right] \right)\delta_{kl},\\
V_{kl} &=  \frac{1}{\sqrt{N}} e^{2 \pi i k l/ N},
\end{align} 
where $[\cdots]$ denotes the floor function which ensures that the eigenvalues, which can be identified with the single-particle momenta $k_j$, are contained inside the first Brillouin zone $(-\pi,\pi]$. Thus it follows that
\begin{align}
\sum_{nm} A^{T}_{mn} c^{\dag}_n c_m =  i \sum_{k\sigma}k c^{\dag}_{k\sigma} c_{k\sigma} =iP.
\end{align}
To compare the Bethe ansatz results of Ref.~\onlinecite{Lieb_uniqueness} with our model we need to choose a gauge in which all hoppings are real. A $\pi$ flux in this gauge corresponds to a Hamiltonian $\mathcal{H}_{-}$ where all the hoppings are equal to $t$ except for the hopping between site $N$ and site $1$, which is equal to $-t$. Under the gauge transformation $G: \mathcal{H}(\phi)\rightarrow\mathcal{H}_-$, the electron creation operator transforms as $G^{-1} c^{\dag}_n G = e^{i\phi n } c^{\dag}_n$. The translation operator $\mathcal{R}_\phi$ is also transformed $\mathcal{R}_{-} = G^{-1} \mathcal{R}_{\phi}G$, and acts on the electron creation operator as 
\begin{equation} 
\mathcal{R}_{-} c^{\dag}_n \mathcal{R}_{-}^{-1} = \begin{cases}
c^{\dag}_n &\text{if } n =1,\ldots,N-1,   \\
-c^{\dag}_n &\text{if } n=N.   \end{cases}
\end{equation}
The order parameter remains the same in first quantization for PBC and aPBC, where a system with PBC and a $\pi$ flux is equivalent to a system with zero flux and aPBC by a large gauge transformation of the basis functions. The sum of the momenta $k$ obtained by the Bethe ansatz for periodic boundary conditions is thus related to the total momentum caculated by $\mathcal{R}$ whereas that same sum for aPBC can be identified with the total momentum calculated by $\mathcal{R}_{\phi}$.

\bibliography{lit}

\end{document}